# NEURAL COMPUTATION AS A TOOL FOR GALAXY CLASSIFICATION : METHODS AND EXAMPLES

O. Lahav[1], A. Naim[1], L. Sodré Jr.[2] and M. C. Storrie-Lombardi[1]

1. Institute of Astronomy, Madingley Rd., Cambridge, CB3 0HA
2. Instituto Astronômico e Geofísico da Universidade de São Paulo,
CEP    CP9638, 01065-970, São Paulo, Brazil

**ABSTRACT.** We apply and compare various Artificial Neural Network (ANN) and other algorithms for automatic morphological classification of galaxies. The ANNs are presented here mathematically, as non-linear extensions of conventional statistical methods in Astronomy. The methods are illustrated using different subsets from the ESO-LV catalogue, for which both machine parameters and human classification are available. The main methods we explore are: (i) Principal Component Analysis (PCA) which tells how independent and informative the input parameters are. (ii) Encoder Neural Network which allows us to find both linear (PCA-like) and non-linear combinations of the input, illustrating an example of unsupervised ANN. (iii) Supervised ANN (using the Backpropagation or Quasi-Newton algorithms) based on a training set for which the human classification is known. Here the output for previously unclassified galaxies can be interpreted as either a continuous (analog) output (e.g. $T$-type) or a Bayesian *a posteriori* probability for each class. Although the ESO-LV parameters are sub-optimal, the success of the ANN in reproducing the human classification is 2 $T$-type units, similar to the degree of agreement between two human experts who classify the same galaxy images on plate material. We also examine the aspects of ANN configurations, reproducibility, scaling of input parameters and redshift information.

## 1 INTRODUCTION

The exponential growth of data in extragalactic Astronomy calls for new approaches to analysis and interpretation. Observations with large ground-based telescopes, automatic measurement machines and satellites have produced large data bases of imaging and spectroscopy of galaxies. However, the advance in producing 'Gigabytes of data' has not been matched by Artificial Intelligence techniques of classification and interpretation. In spite of several attempts (e.g. Murtagh & Heck 1987; Thonnat 1989; Lauberts & Valentijn 1989; Okamura, Kodaira & Watanabe 1984; Spiekermann 1992; Storrie-Lombardi *et al.* 1992; Doi, Fukugita & Okamura 1993; Abraham et al. 1994; Lahav *et al.* 1995; Naim *et al.* 1995b), morphological classification of galaxies still remains a human-intensive process dependent on the eyes of a handful of dedicated individuals.

The motivation for classifying galaxies is two-fold:

(i) RC3-like catalogues for millions galaxies are needed for statistical studies (e.g. correlation functions or density-morphology relation) and as target list of selected type for observational projects, such as Tully-Fisher measurements.

(ii) Classification is important for quantifying the Astrophysics of galaxies, in analogy with the H-R diagram for stars. It allows us to incorporate multi-wavelength and dynamical properties of the galaxies, with the hope that a new 'physical Hubble Sequence' may emerge.

Automated procedures are the only practical way of classifying the enormous amount of data produced by machine scans like those obtained in the Cambridge Automated Plate Measuring (APM) facility and the Sloan Digital Sky Survey (SDSS). The Artificial Neural Network (ANN) method is a novel technique to classify objects which has only little been explored in Astronomy. In a pilot-study (Storrie-Lombardi *et al.* 1992; hereafter SLSS) we have investigated the ANN technique to classify galaxies. Using a Backpropagation algorithm, we have shown that we could reproduce the ESO-LV classification (into 5 classes) at a success rate of 64 % 'perfect match'. More recently, we have shown (Naim *et al.* 1995b; Lahav *et al.* 1995) that ANNs can replicate the human classification of APM-selected galaxies to the same degree of agreement as that between two human experts, 1.8 $T$-type units. This paper provides the theoretical framework and mathematical details of the methods used in these studies.

Other recent applications of ANNs in astronomy include adaptive optics (e.g. Angel et al. 1990), star/galaxy separation (e.g. Odewahn et al. 1991), meteors monitoring (Fraser 1992), and classification of stellar spectra (von Hippel *et al.* 1994). For review of astronomical applications see also Serra-Ricart *et al.* (1993), Miller (1993) and Storrie-Lombardi &

Lahav (1994, 1995). Non-astronomical applications somewhat similar to our problem are speech recognition and identification of hand-written characters. ANNs have several practical advantages compared with traditional techniques. ANN algorithms make no prior assumptions about the statistical distribution of test objects, and invoke no heuristics to define class membership. The algorithms are general-purpose and can be applied to a variety of problems.

Surprisingly, in spite of the wide application of CCD imaging and the theoretical interest in the Hubble Sequence, there is no large uniform data set of galaxy images available. The largest available uniform samples include no more than 200 galaxies (e.g. Kent 1985, Simien & de Vaucoleurs 1986, Kodaira et al. 1986). The recent APM-selected sample of Naim *et al.* (1995a) includes 830 galaxy images. Here we use the ESO-LV data sets, although they are far from being optimal for our problem. They are based on plate material, the galaxies were not classified uniformly by one expert (but by Lauberts, Valentijn and Corwin over a decade) and the machine parameters do not optimally reflect structural parameters like spiral arms which are so apparent to the human eye. However, this is a large data set (more than 5000 galaxies) which includes both machine parameters and human classification. The results presented here should be regarded as a *lower limit* to what can be done with the ANN approach to classification in the future, e.g. with uniform large samples of CCD images which are currently measured (e.g. Madore *et al.* , in preparation; White *et al.* , in preparation).

In this paper we shall also address briefly astrophysical implications of our ANN results. One open question is whether galaxies were formed in a self-similar way, or in a way which mainly depends on the their total mass or potential well. For example, Simien & de Vaucoulerus (1986) showed a tight correlation between the disk-to-bulge ratio (a distance independent property) and the Hubble type, while Meisles & Ostriker (1984) argued that the absolute luminosity of the spheroidal component (a distance dependent property) is the major parameter determining the Hubble Sequence. We shall examine this question using ESO-LV diameters.

The structure of the paper is as follows. As the ANN methods are general and are currently scattered in the ANN literature (e.g. in journals of Engineering and Biology), we present them mathematically in Appendices A (Principal Component Analysis and its non-linear extensions), B (Backpropagation and Quasi-Newton minimization algorithms) and C (Bayesian classification, Wiener filtering and weight decay). The main text of the paper gives examples of applications of these methods to the ESO-LV galaxies. Following a general Introduction (§1), §2 presents the ESO-LV parameters. §3 illustrates the use of Principal Component Analysis, while §4 presents a variety of applications of supervised non-linear ANNs. Future work is discussed in §5.

## 2 THE DATA SETS

Here we illustrate the method using ESO-LV galaxies (Lauberts & Valentijn 1989; hereafter LV89) at high Galactic latitude ($|b| > 30^o$). We shall consider several samples. There are three aspects in defining the samples for training by the ANN : the sample selection (e.g. by apparent diameter), the galaxy machine parameters used, and the binning into galaxy classes.

The first sample, composed of 13 galaxy parameters, hereafter called $P13$, is the same as we used in SLSS, i.e. of galaxies with visual diameters larger than 1 arcmin (the claimed completeness of the ESO-LV catalogue). Only galaxies with morphological classification performed by visual examination of the galaxy image were considered in our analysis. We use the 13 catalogue parameters shown in Table 1 of SLSS to describe each galaxy. Briefly they are : (1) the average blue minus red colour, (2) the exponent in the generalized de Vacouleurs law in the blue, (3) log of the ratio of diameters which include 80 % and 50 % of the blue light, (4) an indicator of the degree of asymmetry of the galaxy image, (5) the central blue surface brightness, (6) log of the ratio of minor to major diameters, (7) error in ellipse fit to blue isophotes, (8) gradient of the blue surface brightness profile at half-light radius, (9) log of the ratio of the blue 26 mag/arsec$^2$ diameter and the half-light diameter, (10) the exponent in the generalized de Vacouleurs law in the red, (11) average blue surface brightness within 10 arcsec diameter circular aperture, (12) blue surface brightness at half-light radius, (13) red surface brightness at half-light radius. These 13 parameters were chosen because they are almost distance-independent, and they are very similar to those used by LV89 to perform the automated classification presented in the ESO-LV catalogue. This allows us to compare meaningfully the success rate of the classifications provided by our ANN with ESO-LV. After selecting only galaxies with all 13 parameters available, our final data set has 5217 galaxies. We then randomly divided these galaxies in two independent sets of 1700 and 3517 objects for training and testing. We have also normalized our input data between 0 and 1 by using the minimum and maximum values of each parameter (and also have tried normalizing by the variance). We have grouped the ESO-LV catalogue sub-classes in three ways: (i) by keeping the original range of classes $-5.0 \leq T \leq 10.0$ where $T$ is the coded type; (ii) by binning into 5 major classes (as in SLSS): E ($-5.0 \leq T < -2.5$; 466 galaxies); S0 ($-2.5 \leq T < 0.5$; 851 galaxies); Sa+Sb ($0.5 \leq T < 4.5$; 2403 galaxies); Sc+Sd ($4.5 \leq T < 8.5$; 1132 galaxies); and Irr ($8.5 \leq T \leq 10.0$; 365 galaxies), and by splitting into two classes : early type, (E+S0, $T < 0.5$, 1317 galaxies ) and late type ($T \geq 0.5$, 3900 galaxies).

The second sample, hereafter called $D7$, is also extracted from ESO-LV. It includes galaxies larger than 2 arcmin (as defined by the old ESO sample) which also have redshift information, and information of 7 diameters. $D_e$, $D_{70}$, $D_{80}$ and $D_{90}$ are the major diameters of ellipse at 50 %, 70 %, 80 % and 90 % total $B$ light, while $D_{25}$, $D_{26}$ and $D_{27}$ are the major diameters of ellipse at $B$ surface brightness of of 25, 26 and 27 $mag/arcsec^2$. We then converted them into metric diameters using their redshift. This sample includes 791 galaxies, which were mainly classified by one expert, H. Corwin.

# 3 HOW INFORMATIVE ARE THE INPUT PARAMETERS ?

A key question when providing an ANN with an input is how many input parameters to present, and how to compress them in an efficient and informative way. There is of course a trade-off between keeping the number of parameters small and the amount of information presented.

### 3.1 Standard PCA

Principal Component Analysis (PCA) is a widely used method which allows to judge how many independent parameters are needed, by looking at directions along which the variance is maximal. The formulation of PCA is given in Appendix A.1. It is worth emphasizing that PCA is only meaningful for linear parameters (or 'the nearest to linear', e.g. by taking log of the original variables), and may suffer from scaling problems. Nevertheless, it is a useful tool for reducing the dimensionality of the input parameter space. In the context of this paper it can be viewed as a data compression techinque for the input to the ANN, as well as an 'unsupervised method' for exploring the parameter space.

We begin by applying the method to the log of the 7 metric diameters given in the $D7$ sample (of 791 galaxies), each scaled to have zero-mean. We do not normalize here by the variance of each variable, as they all have the same metric, and we wish to represent their relative values. Not too surprisingly, the correlation matrix indicates strong correlation between the log-diameters. Table 1 gives the eigen-values and the eigen-vectors for the log-diameters. 95 % of the variance is in the first principal component (which is found to be approximately the average of the 7 log-diameters). However, as we show in §4.4 , using the ANN, it is *not* sufficient to use just the first principal component to represent the 7-dimensional data space for classification.

We then applied PCA to the 13 distance-independent parameters of the $P13$ sample of 5217 galaxies, with the parameters normalized to have zero-mean and unit-variance, as here the parameters are made of 'apples and oranges'. Indeed, one should be more cautious about applying PCA to a set of parameters which are of different character, and may well be non-linear. However, the results give some insight into the amount of useful information in this parameter space. Tables 2a and 2b give the 13 eigen-values and the eigen-vectors corresponding to the largest 3 eigen-vectors. We find that the first 7 linear combinations give 90 % of the variance.

The projection of the 13 parameters on the first and second Principal Components is shown in Figures 1a,b. Although the distribution of all galaxies looks like a fuzzy cloud, the different morphological types actually occupy distinct regions in this new parameter space. We see that even E and S0 galaxies can actually be separated. This plot illustrates how PCA could compress a 13-dimensional parameter space into a 2-dimensional space. Although the physical meaning of the new space is not easy to interpret, it allows to segregate different classes of objects.

### 3.2 Encoder and Neural PCA

Generally, a multi-layer ANN consists of nodes (analogous to human neurons) arranged in a series of layers. The nodes in a given layer are fully connected to the nodes in the next layer. The free parameters of the ANN are the weights $w_{ij}$ which are determined by least-squares of the difference between the input and the desired output, the so-called 'cost

function':

$$E = \frac{1}{2} \langle \sum_k (o_k - d_k)^2 \rangle, \qquad (1)$$

where the sum is over the components of the vector ($k = 1, M$) and the average is over the galaxies. Layers between the input and the output layers are called 'hidden layers' and allow non-linear mapping. The least-squares minimization can be done by a variety of efficient algorithms, e.g. Backpropagation and Quasi-Newton, which are described in detail in Appendix B.

We begin by demonstrating an encoder network in which the desired output is the input itself, as explained in detail in Appendix A.2. Figure 2 shows an $M : M' : M$ network configuration, where $M'$ is the number of 'neck' units (in the 'hidden layer'), or number of linear combinations in the PCA language. While a linear network in this configuration simply reproduces a standard PCA, a non-linear transfer function can allow 'non-linear PCA'.

We now apply a non-linear encoder network, with a sigmoid threshold function. In Figure 3 we plot the cost function (calculated over the 5217 ESO-LV galaxies) vs. the number of hidden units $M'$. Clearly when the input 13 parameters are uncorrelated, we shall need 13 hidden units to fully recover the 13 parameters at the output layer. If, on the other hand, the 13 parameters are identical, then one hidden unit will be sufficient. The figure shows that the cost function drops exponentially as a function of the number of hidden units. This behaviour may serve for guidance in selecting the number of hidden units for the classification network (see below). Serra-Ricart *et al.* (1993) have developed this unsupervised approach further, illustrating for our P13 data set that a non-linear encoder can identify classes in this data set much better than a standard PCA. Other algorithms for neural PCA such as Oja's rule are discussed in Appendix A.3 .

## 4 SUPERVISED CLASSIFICATION OF GALAXIES WITH ANNs

We now apply 'standard' supervised ANN for classifying the ESO-LV galaxies. In the 'training' process, the input vectors, containing the galaxy parameters, are presented to the network. The weights ('free parameters') $w_{ij}$ are computed by least-squares minimization with the Backpropagation or Quasi-Newton algorithms (explained in detail in Appendix B). The ANN is then ready to handle new unclassified data for which only the machine parameters are available. We shall present 3 different net configurations: (i) a single output ('analog') network, (ii) a two-class classifier, and (iii) a 5 -class classifier. We wish to emphasize that *supervised* ANNs do not produce an 'objective' unique classification. Supervised networks replicate the choices of their trainer - a network trained according to the classification made by Hubble or de Vaucouleurs will classify new data in a manner similar to the original expert.

### 4.1 Single Continuous Output

Although the galaxy morphology is probably a continuous sequence (Hubble 1936), human experts provide us with a 'true answer' usually given in quantized units, to a first decimal point, e.g. $T = 5.3$. It is to our benefit that the single output configuration of the network can approximate a one-dimensional continuous sequence.

It is common in Astronomy to fit a model with several free parameters to the observations. This regression is usually done by means of $\chi^2$ minimization. A simple example of a model is a polynomial with the coefficients as the free parameters. Consider now the specific problem of morphological classification of galaxies. If the type is $T$ (e.g. on the numerical system $[-5, 10]$), and we have a set of parameters $\mathbf{x}$ (e.g. isophotal diameters and colours) then we would like to find the free parameters $\mathbf{w}$ ('weights') such that the cost function

$$E = \frac{1}{2} \sum_i [T_i - f(\mathbf{w}, \mathbf{x}_i)]^2, \qquad (2)$$

is minimized. The function $f(\mathbf{w}, \mathbf{x})$ is the 'network'. Commonly $f$ is written in terms of the variable

$$z = \sum_k w_k x_k, \qquad (3)$$

where the sum here is over the input parameters to each node. A 'linear network' has $f(z) = z$, while a non-linear threshold function could be a sigmoid $f(z) = 1/[1 + \exp(-z)]$ or $f(z) = \tanh(z)$. Another element of non-linearity is provided by the the 'hidden-layers'. The 'hidden-layers' allow curved boundaries around clouds of data points in the parameter space. A typical configuration with one 'hidden-layer' and a single output is shown in Figure 4. While in most computational problems we only have 10-1000 nodes, in the brain there are $\sim 10^{10}$ neurons, each with $\sim 10^4$ connections. We do not regard of course our simple ANN algorithm as a model for the human brain, but rather as a non-linear statistical method.

The determination of many free parameters, the weights $w_i$'s in our case, might be unstable. It is therefore convenient to regularise the weights, e.g. by preventing them from growing too much. In the ANN literature this is called 'weight decay'. This approach is analogous to Maximum Entropy, and can be justified by Bayesian arguments, with the regularising function acting as the prior in the weight space. One possibility is to add a quadratic prior to the cost function and to minimize

$$E_{tot} = \alpha E_w + \beta E_D, \qquad (4)$$

where $E_D$ is our usual cost function, based on the data (e.g. eq. 2) and

$$E_w = \frac{1}{2} \sum_{i=1}^{Q} w_i^2 \qquad (5)$$

is the chosen regularising function, where $Q$ is the total number of weights. The coefficients $\alpha$ and $\beta$ can be viewed as 'Lagrange multipliers'. While sometime they are specified ad-hoc, it is possible to evaluate them 'objectively' by Bayesian arguments in the weight-space. We discuss this procedure in detail in Appendix C.3 .

To illustrate the above ideas, we built a network with configuration 13:3:1, resulting in 46 free weights (including the 'bias' node, which represents an additive constant). In the training process the network was presented with 13 parameters (from sample P13) for each galaxy, using a subset of 1700 galaxies. Both the input parameters and the 'true' answer $T$-type (in the range $-5 \leq T \leq 10$) were scaled to the range $[0, 1]$, so all the weights were treated on equal footing in the regularisation process. The transfer function used was a sigmoid. By the procedure outlined in Appendix C.3 we found the weight decay regularization coefficient to be $\frac{\alpha}{\beta} = 0.001$. We then applied least-square minimization using a Quasi-Newton method (as implemented in a code kindly provided to us by B.D. Ripley).

As in other optimization problems, it is crucial to decide when to stop the minimization. One approach is to stop when the cost function drops below a certain value, or changes little between successive iterations. However, in particular when the sample size is small (relative to the number of weights) this may result in 'over-fitting' ('memorising') of the data (including the noise). Usually, the cost function with respect to the training set shows monotonic decline, and it is difficult to define a minimum for stopping. Instead, we calculate at each iteration the cost function for the testing set (with the weights derived of course from the training set). In this way we monitor the ability of the ANN to 'generalize' its choice of weights to data it was not trained on. Usually the cost function with respect to the testing set decreases to a minimum and then increases, so it is easy to decide where to stop according to this minimum.

Once the training phase was completed, we presented the network with a testing set (of 1700 galaxies), but for which a human classification is known. On a Sun Sparc workstation the training of the network on 1700 galaxies takes about 1 min (CPU), while testing on a sample of similar size takes only 1 sec (CPU).

Figure 5 shows the network type $T_{net}$ versus the ESO-LV human classification $T_{eso}$. The Spearman rank-order correlation coefficient is $r_s = 0.83$. As another way of quantifying the network performance we calculate the variance between the network and the ESO-LV type over the number of galaxies $N_{gal}$ :

$$\sigma^2 = \frac{1}{N_{gal}} \sum (T_{net} - T_{eso})^2 \qquad (6)$$

and we find $\sigma \approx 2.0$ $T$-units. By a similar statistics we can compare the run of similar network configuration which start the minimization with different random weights. Figure 6 shows the results of the two runs. The scatter between two runs is much smaller than that in Figure 5. Here the Spearman coefficient is $r_s = 0.98$ and typical 'reproducibility' scatter is $\sigma \approx 0.6$ $T$-units. In Figure 6 we note a 'break' in the transition from early type ($T < 0$) to late type ($T > 0$). It may be that the non-linearity of the network was not sufficient to fit both classes by the same weights (i.e. in each minimization the net finds a different compromise of weights to satisfy both early and late type galaxies), or that the quality of parameters for early and late types is different.

In the study of the blue APM images (Naim et al. 1995b, Lahav et al. 1995) we have shown that ANNs can replicate the classification by a human expert to the same degree of agreement as that between two human experts, to within 1.8 $T$-units (based on a comparative study where the same images were classified by 6 experts independently). The ESO-LV data give a slightly weaker result, 2 $T$-units, probably due to the parameters being less informative, although they include blue minus red colour as a parameter, which is lacking in the APM sample.

### 4.2 Two-Class (E, S) Classifier

In a network with multiple outputs, the output vector can be interpreted in a probabilistic way. The $j-th$ component of this vector can be viewed as the probability for class $j$ given the input parameters, $P(C_j|\mathbf{x})$. In fact, it can be proved theoretically (Appendix C.1) that the output of an ideal ANN is indeed a Bayesian *a posteriori* probability. Moreover, as our experiments confirm, the sum of the output vector components is $\sum_k o_k \approx 1$, as expected for a probabilistic classifier. It is worth noting that, unlike discrete classification of hand-written characters, galaxies form a continuous sequence. Hence the combination of probabilities assigned to different 'eigen-classes' may reflect an intermediate class. The 'most likely class' can be defined as the class associated with the largest output component. Here we do not include weight decay, as when included the interpretation of the output is not anymore strictly Bayesian (see Appendix C.1).

To classify into early type ($-5 \leq T \leq 0.5$) and late type ($0.5 < T \leq 10$) galaxies we have used a Backpropagation algorithm (Appendix B) and a network configuration of 13:10:2 with a tanh threshold function, a learning coefficient $\eta = 0.01$, a momentum coefficient $\alpha = 0.9$ (see Eq. B6). The network was trained on 1700 ESO-LV galaxies, and was tested on the remaining 3517 galaxies. Of the 898 galaxies classified as early type by ESO-LV, 681 were classified as such by the network, while 217 were classified as late type. Of the 2619 galaxies classified as late type by ESO-LV, 2471 were classified as such by the network, while 148 were classified as early type type. This means a success rate of 90%.

Breaking down the early class into ellipticals and lenticulars (S0s) demonstrates that the vast majority of the variance is in the classification of S0s. Of the 311 galaxies classified as ellipticals ($-5 \leq T < -2.5$) by ESO-LV the ANN agreed on 94% of them and disagreed for only 6%. On the other hand, of the 587 galaxies classified as S0 ($-2.5 \leq T < 0.5$) by ESO-LV, the ANN agreed only for 66% and disagreed for 34%. This is yet another indication that the S0s form a 'transition class' along the Hubble sequence.

### 4.3 Five-Class Classifier

This is essentially the network we presented in SLSS. The input layer consists of the 13 parameters and the output layer consists of the 5 classes described in section 2. The configuration used was (13;13,5) with sigmoid as our nonlinear transfer function. The learning and momentum coefficients were kept constant at $\eta = 0.5$ and $\alpha = 0.2$, for all layers.

During training (using 1700 ESO-LV galaxies of the P13 sample), the ANN compared the output of these five nodes to the visual classification decisions of LV89. We then tested the ANN against the remaining 3517 galaxies of the P13 sample. Morphological classification was performed by assigning the galaxy to the class corresponding to the maximal output component. Further experiments we carried out with a variety of network configurations showed little effects of the number of hidden units and layers, the epoch, the size of training set, the number of iterations, and the learning and momentum coefficients.

Our main result, shown in table 2 of SLSS still stands. The percentage of galaxies correctly classified was 64 % (and 96 % to within nearest class; If either the first or the second highest outputs are considered in the comparison with the visual classification, the success rate is 90 %). On the other hand, a simple Bayesian classifier we constructed (assuming Gaussian multi-variate function, see Appendix C.1, eqs. C1 and C2) only gave 56 %. This is the same success rate reported by LV89 by their linear classifier. This clearly shows that non-linear ANNs can be superior to linear classifiers, and that the classifier itself is of great importance, not only the parameters.

### 4.4 PCA data compression as input to ANNs

In this section we address the question how many Principal Components are needed to recover the same classification achieved with ANN using the full input data. To illustrate this point we use the D7 sample of 791 ESO-LV galaxies, where the input parameters are the log of 7 *metric* diameters. The network architecture is 7:3:1, with both input and output scaled to $[0, 1]$, and using the Quasi-Newton algorithm, with weight decay coefficient $\frac{\alpha}{\beta} = 0.001$. Training was done on 600 galaxies, and testing on the remaining 191 galaxies. The resulting rms scatter (eq. 6) evaluated over several runs is $\sigma = 2.2$. Using as input only the first Principal Component, which was derived in section 3.1 and accounts for 95% of the variance, we find a much larger scatter, $\sigma = 3.6$. Only when the first 3 PC's are used, does one recover the scatter achieved by using all 7 diameters. This shows that the fractional variance on its own is not sufficient to tell us how many PC's are needed for classification.

The failure of the first Principal Component to recover on its own the classification might be due to non-linearity in the data, the effect of noise on the deduction of Principal Components, or the fact that classification requires more information than that given just by the maximal variance (i.e. the second moment of the distribution function). We note that commonly the fractional variance of the eigen-values is used as the sole criterion in compressing the data prior to applying a classification procedure (e.g. in deriving the 'concentration parameter', of Okamura *et al.* 1984). However, this criterion may underestimate the importance of the minor Principal Components. It may well be that classification can be improved by using more principal components. Furthermore, our experience shows that in some cases minor Principal Components are more important than major Principal Components.

### 4.5 Scaled parameters vs. absolute parameters

So far, in this paper as well as in our previous studies (SLSS, Naim et al. 1995b, Lahav et al. 1995), we have not used the distance (as estimated from the redshift) to the galaxies. Our input parameters were always scaled, such that they were distance independent. In a sense, we have assumed that what matters in classification are the relative properties of the galaxies. e.g., that two ellipticals with high and low absolute luminosities will be classified as the same type if one is a scaled down (or up) version of the other.

The astrophysical question whether galaxies were formed in a self-similar way, or in a way which mainly depends on the their total mass or potential well is still open. For example, Simien & de Vaucouleurs (1986) showed a tight correlation between the disk-to-bulge ratio (a distance independent property) and the Hubble type, while Meisels & Ostriker (1984) argued that the absolute luminosity of the spheroidal component (a distance dependent property) is the major parameter determining the Hubble Sequence.

To test this question we have used the D7 data as described in §4.4 and fed the ANN with the log of ratio of 6 diameters to the half-light diameter. The resulting scatter was larger, $\sigma = 2.4$, compared with scatter of 2.2 when all 7 metric diameters were presented. Our tentative conclusion is that absolute parameters are not much more informative than the scaled properties. However, the quality of the data and the parameters used (diameters) are not sufficient to prove the theoretical prejudice some may have that only scaled (self-similar) properties control the fate of a galaxy along the Hubble Sequence.

## 5 DISCUSSION

In this paper we have attempted to de-mystify ANNs by showing how they generalize other statistical methods commonly used in Astronomy and other fields. The methods were illustrated using the ESO-LV galaxy data, showing that ANNs can replicate successfully human classification. These results for ESO-LV are in accord with our results for the APM sample of 830 galaxy images (Lahav et al. 1995, Naim et al. 1995b): an ANN can replicate the classification by a human expert to within 2 $T$-type units, similar to the scatter between two human experts.

ANNs are sometime considered as being esoteric methods. Questions commonly asked by 'Neuro-sceptics' are: (i) Could we understand what the ANNs are doing, or are they just 'black boxes'? (ii) If one has already selected 'good parameters', does it matter what classifier is to be used ? We have shown that the ANNs approach should be viewed as a general statistical framework. Some special cases of ANNs are statistics we are all familiar with. However, the ANNs can do better, by allowing non-linearity. There is of course freedom in choosing what kind of 'non-linearity' to apply, but sensible choices show that significant improvement can be achieved over the linear approaches. For cosmologists, there is an analogy here with $N$-body simulations of gravitational systems. Linear theory is reasonably well understood, but is not sufficient to describe complicated dynamics. One needs to use then numerical simulations, producing results which are not always understood by intuition or by analytic methods. However, one can verify what is happening by considering simple cases (e.g. the spherical infall model) to gain confidence in what the simulations give. Our approach to the ANNs is similar.

This paper does not cover of course all possible approaches to classification. For example, as described in Appendix C.2, one can use just a linear network (in which the weights effectively act like a Wiener filter), but modify the input parameters to be non-linear (in a somewhat ad-hoc way). In some case such networks can do as well as the non-linear ANNs. Another important issue, not discussed here, is how to handle noisy data.

An even more challenging task is to devise 'unsupervised' algorithms, where there is no external 'teacher', and the data speak for themselves. Such methods could be either 'cooperative' (e.g. PCA, non-linear encoder, or the Kohonen 1989 self-organizing map) or 'competitive' (e.g. cluster analysis). For preliminary applications of unsupervised methods to galaxy classification see Naim (1995). These unsupervised algorithms may well explore new features in the data set which were previously ignored by the human experts.

On the more astrophysical side, the goals are to incorporate dynamical properties of galaxies (e.g. circular velocities) and multi-wavelength data (from radio to the UV). The hope is that these methods will help defining a new physical space of galaxies, in analogy with the H-R diagram for stars.

**Acknowledgments**. We thank W. Fitzgerald, S. Folkes, J. Hertz, M. Irwin, J. Lasenby, D. Lynden-Bell, D. MacKay, B. Ripley and T. von Hippel for stimulating discussions. OL acknowledges the hospitality of the Hebrew University and the Weizmann Institute, where parts of this paper were written, AN thanks an Overseas Research Studentship and an Isaac Newton Studentship, LSJ acknowledges FAPESP and CNPq for financial support, and MCSL thanks the Sheepshanks Fund and the Harrison-Watson Foundation of Clare College for financial support.

# APPENDIX A: PRINCIPAL COMPONENT ANALYSIS (PCA) AND ANN

## A.1 Standard PCA

A pattern can be thought of as being characterized by a point in an $M$-dimensional parameter space. One may wish a more compact data description, where each pattern is described by $M'$ quantities, with $M' \ll M$. This can be accomplished by Principal Component Analysis (PCA), a well known statistical tool, commonly used in Astronomy (e.g. Murtagh & Heck 1987 and references therein). The PCA method is also known in the literature as Karhunen-Loéve or Hotelling transform, and is closely related to the technique of Singular Value Decomposition. By identifying the *linear* combination of input parameters with maximum variance, PCA finds $M'$ variables ('Principal Components') that can be most effectively used to characterize the inputs.

The first Principal Component is taken to be along the direction in the $M$-dimensional input parameter space with the maximum variance. More generally, the $k$-th component is taken along the maximum variance direction in the subspace perpendicular to the subspace spanned by the first $(k-1)$ Principal Components. It is convenient to apply PCA to data already standardized, e.g. transformed to zero-mean and unit-variance. However, while this scaling is appropriate for data composed of 'apples and oranges' as in the present paper for the 13 ESO-LV parameters, in other problems such as the 7 ESO-LV diameters and spectral analysis of quasars and galaxies (cf. Francis et al. 1992, Lahav 1995) it is more sensible not to divide by the variance of each channel (over an ensemble of objects), in order to keep the relative strength of the lines.

The formulation of standard PCA is as follows. Consider a set of $N$ objects ($i = 1, N$), each with $M$ parameters($j = 1, M$). If $r_{ij}$ are the original measurements, we construct normalized properties as follows:

$$x_{ij} = \frac{r_{ij} - \bar{r}_j}{s_j}, \qquad (A1)$$

where $\bar{r}_j = \frac{1}{N} \sum_{i=1}^{N} r_{ij}$ is the mean, and $s_j^2 = \frac{1}{N} \sum_{i=1}^{N} (r_{ij} - \bar{r}_j)^2$ is the variance. We then construct a correlation matrix

$$C_{jk} = \frac{1}{N} \sum_{i=1}^{N} x_{ij} x_{ik} \qquad (A2)$$

It can be shown that the axis along which the variance is maximal is the eigen-vector $\mathbf{u_1}$ of the matrix equation

$$C\mathbf{u_1} = \lambda_1 \mathbf{u_1}, \quad (A3)$$

where the $\lambda_1$ is the largest eigen-value, which is in fact the variance along the new axis. The other principal axes and eigen-vectors obey similar equation. It is convenient to sort them in decreasing order, and to quantify the fractional variance by $\lambda_\alpha / \sum_\alpha \lambda_\alpha$. It is also convenient to re-normalize each component by $\sqrt{\lambda_\alpha}$, to give unit-variance along the new axis. We note that the weakness of PCA is that it assumes linearity and also depends on the way the variables are scaled. In contrast, ANNs generally allow non-linearity.

### A.2 Principal Components from Encoder

PCA is in fact an example of 'unsupervised learning', in which an algorithm or a 'linear network' discovers for itself features and patterns (see e.g. Hertz *et al.* 1991 for review). A simple net configuration $M : M' : M$ (see Fig. 2) with linear transfer functions allows finding $M'$ linear combinations of the original $M$ parameters. The idea is to force the output layer to reproduce the input layer, by least-squares minimization (e.g. using the Backpropagation algorithm, see Appendix B). If the number of 'neck units' $M'$ equals $M$ then the output will exactly reproduce the input. However, if $M' < M$, the net will find, after minimization, the optimal linear combination. By changing the threshold function from linear to non-linear (e.g. a sigmoid) one can allow 'non-linear PCA'. Some authors (e.g. Geva & Sitte 1992; Serra-Ricat et al. 1993) advocate a configuration of $M : 2M + 1 : M' : 2M + 1 : M$ to get optimal reconstruction of non-linear shapes, e.g. a circle.

### A.3 Neural PCA: Oja's Neural Network

One interesting aspect of ANN theory is that a very simple artificial neuron can be trained to extract the first Principal Component of the input parameters (Oja 1982). Consider an artificial neuron which receives a set of $n$ scalar-valued inputs $x_1, ..., x_n$ through $n$ connections with coupling strengths (weights) $w_1, ..., w_n$ and produces an output $Y$:

$$Y = \sum_{i=1}^{n} w_i x_i. \quad (A4)$$

During the training of this neuron, the weights $w_i$ can be changed after the presentation of a pattern according to the Hebbian rule (Hebb 1949),

$$\Delta w_i = \eta Y x_i \qquad (A5)$$

where $\eta$ controls the rate of learning. Oja (1982) generalized this rule by incorporating a normalization

$$w_i(t+1) = \frac{w_i(t) + \eta Y x_i}{\{\sum_{j=1}^{n}[w_j(t) + \eta Y x_j]^2\}^{\frac{1}{2}}}, \qquad (A6)$$

where $\eta$ is the 'learning coefficient'. Expanding this expression as a power series in $\eta$ and retaining only the first order term yields the learning equation known as Oja's rule:

$$w_i(t+1) = w_i(t) + \eta Y(t)[x_i - Y(t)w_i(t)] \qquad (A7)$$

Oja's rule, after training, chooses the direction of the weights vector $\mathbf{w}$ to lie in the maximal eigenvector direction of the correlation matrix $\langle \mathbf{x}\mathbf{x}^\mathbf{T} \rangle$ (assuming zero-mean data, and using here matrix multiplication notation; $\mathbf{x}^T$ being the transposed vector ). Moreover, this turns out to be also the direction which maximizes the variance of the output $\langle Y^2 \rangle = \mathbf{w}^\mathbf{T} \langle \mathbf{x}\mathbf{x}^\mathbf{T} \rangle \mathbf{w}$ (see e.g. Hertz et al. 1991). Oja (1982) also showed that, after training, the normalization $\sum_{i=1}^{n} w_i^2$ tends to be bounded and close to one. Other rules to extract the first and higher Principal Components have been proposed e.g. by Sanger (1989) and Oja (1992). While these learning rules give insight to the link between PCA and ANN, it is easier in practice to extract the Principal Components by the standard method (Appendix A.1) or by a linear encoder (Appendix A.2).

## APPENDIX B: MINIMIZATION ALGORITHMS

### B.1 The Backpropagation method

The Backpropagation algorithm has been re-invented several times (e.g. Werbos 1974; Parker 1985; Rumelhart, Hinton & Williams 1986) and is one of the most popular ANN algorithms. A typical configuration is shown in Fig. 4. For a given network architecture the first step is the 'training' of the ANN. In this step the weights $w_{ij}$'s (the 'free parameters') are determined by minimizing 'least-squares'. The novel aspect of Backpropagation is the way this minimization is done, using the chain rule (gradient descent).

Each node (except the input nodes) receives the output of all nodes in the previous layer and produces its own output, which then feeds the nodes in the next layer. A node at layer $s$ calculates a linear combination over the input $x_i^{(s-1)}$ from the previous layer $s-1$ according to

$$I_j^{(s)} = \sum_{i=0}^{n} w_{ij}^{(s)} x_i^{(s-1)} \qquad (B1)$$

where the $w_{ij}$'s are the weights associated with that node. Commonly one takes $x_0 = 1$, with $w_{0j}$ playing the role of a 'bias' or DC level. The node then fires a signal

$$x_j^{(s)} = f(z), \qquad (B2)$$

where $z$ here stands for $I_j^{(s)}$, and $f$ is a non-linear transfer function usually of the sigmoid form

$$f(z) = 1/[1 + \exp(-z)] \qquad (B3)$$

in the interval [0,1], or

$$f(z) = \tanh(z) \qquad (B4)$$

in the interval [-1,1].

For each object (pattern) in the training set, the network compares its output vector in the 'classification space' **o** to the desired vector **d** determined by the 'true answer' (e.g. as given by a human expert). For example, the elements of the vector **d** can be defined as zeros except for one element set to 1 corresponding to an actual class, e.g. we define **d** = $(1, 0, 0, 0, 0)$ for Elliptical galaxies.

The comparison is done in terms of a cost function, usually of the form

$$E = \frac{1}{2} \sum_k (o_k - d_k)^2 , \qquad (B5)$$

where the sum is over the components of the vectors. This cost function, averaged over all the training galaxies presented to the ANN is minimized with respect to free parameters, the weights $w_{ij}$. The weights are updated by gradient descent *backwards* (hence the name Backpropagation) from the output layer to one or more hidden layers, by a small change in each time step,

$$\Delta w_{ij}(t+1) = -\eta \frac{\partial E}{\partial w_{ij}} + \alpha \Delta w_{ij}(t), \qquad (B6)$$

where the 'learning coefficient' $\eta$ and the 'momentum' $\alpha$ are 'knobs' which control the rate of learning and the inertia from the previous iteration respectively (see e.g. Hertz *et al.* 1991).

The elegance of the Backpropagation algorithm is in the way the derivative is evaluated. Let us consider the case of a sigmoid output

$$o_j^{(s)} = x_j^{(s)} = f(I_j^{(s)}) \qquad (B7)$$

where

$$f(z) = \frac{1}{1 + e^{-z}}. \qquad (B8)$$

In this case note that $f' = f(1 - f)$ and the derivative can be written as:

$$\frac{\partial E}{\partial w_{ij}} = o_i^{(s-1)} o_j^{(s)} (1 - o_j^{(s)}) \beta_j, \qquad (B9)$$

where

$$\beta_j = (o_j - d_j) \qquad (B10)$$

for nodes in the *output* layer, and

$$\beta_j = \sum_k w_{jk} o_k (1 - o_k) \beta_k \qquad (B11)$$

for nodes in *hidden* layers (the sum is over $k$ nodes in the layer *above* node $j$).

The 'hidden layers' allow curved boundaries around clouds of data points in the parameter space in a non-parametric way. The interpretation of the output depends on the network configuration. For example, a single output node provides a continuous output (e.g. predicting the $T$-type as in §4.1 or the luminosity of a galaxy), while several output nodes can be used to assign probabilities to different classes (e.g. 5 morphological types of galaxies), as explained in Appendix C.

**B.2 The Quasi-Newton method**

There are methods, other than Backpropagation, for minimizing the non-linear function eq. (B5). A more efficient method is Quasi-Newton. In short, the cost function $E(\mathbf{w})$ in terms of the weights vector $\mathbf{w}$ is expanded about a current value $\mathbf{w_0}$ :

$$E(\mathbf{w}) = E(\mathbf{w_0}) + (\mathbf{w} - \mathbf{w_0}) \nabla E(\mathbf{w_0}) + \frac{1}{2}(\mathbf{w} - \mathbf{w_0}) \cdot \mathbf{H} \cdot (\mathbf{w} - \mathbf{w_0}) + ..., \qquad (B12)$$

where $\mathbf{H}$ is the Hessain with elements $H_{ij} = \frac{\partial^2 E}{\partial w_i \partial w_j}$ evaluated at $\mathbf{w_0}$. The minimum approximately occurs at

$$\nabla E(\mathbf{w}) \approx \nabla E(\mathbf{w_0}) + \mathbf{H} \cdot (\mathbf{w} - \mathbf{w_0}) = 0. \qquad (B13)$$

Hence an estimation for the optimal weights vector is at

$$\mathbf{w} = \mathbf{w_0} - \mathbf{H}^{-1} \nabla E(\mathbf{w_0}). \qquad (B14)$$

In the standard Newton's method a previous estimate of $\mathbf{w}$ is used as the new $\mathbf{w_0}$. Calculating the Hessian exactly is expensive computationally, and in the quasi-Newton method an iterative approximation is used for the inverse of the Hessian (e.g. Press et al. 1992; Hertz et al. 1991).

## APPENDIX C: RELATIONS BETWEEN ANNs AND OTHER CLASSIFIERS

### C.1 Bayesian Classification and probabilities

A classifier can be formulated from first principles according to Bayes theorem:

$$P(T_j|\mathbf{x}) = \frac{P(\mathbf{x}|T_j) \, P(T_j)}{\sum_k P(\mathbf{x}|T_k) \, P(T_k)} \qquad (C1)$$

i.e. the *a posteriori* probability for a class $T_j$ given the parameters vector $\mathbf{x}$ is proportional to the probability for data given a class (as can be deduced from a training set) times the *prior* probability for a class (as can be evaluated from the frequency of classes in the training set). However, applying eq. (C1) requires parameterization of the probabilities involved. It is common, although not always adequate, to use multivariate Gaussian:

$$P(\mathbf{x}|T_j) = (2\pi)^{-M/2} \, |C_j|^{-1/2} \, \exp[-\frac{1}{2}\mathbf{x} \, C_j^{-1} \, \mathbf{x}^T], \qquad (C2)$$

where $\mathbf{x}$ is of dimension $M$ and here has zero-mean, $\mathbf{x}^T$ is its transposed vector, and $C_j = \langle \mathbf{x} \, \mathbf{x}^T \rangle_j$ is the covariance matrix *per class j*. This matrix is similar to the one used in the PCA (Appendix A.1) for *all the classes.* As in PCA, the matrix $C_j$ can be diagonalized, hence simplifying eq. (C2).

It can be shown that certain ANN configurations behave like Bayesian classifiers, i.e. the output nodes produce Bayesian *a posteriori* probabilities (see e.g. Gish 1990; Richard & Lippmann 1991), although it does not implement Bayes theorem directly. To illustrate this important property of the networks we follow Gish (1990) for a simple heuristic example. Let the network's single output be written as $f(x,w)$ where $x$ stands for the input parameters and $w$ stands for the weights (more generally these quantities are vectors). We consider a two class problem for which the desired output of the network is 1 if $x$ is in class $T_1$ and 0 if it is in class $T_2$. The cost function over all objects $N$ is then (cf. eq. B5)

$$E = \frac{1}{N}\{\sum_{x \in T_1}[f(x,w) - 1]^2 \ + \ \sum_{x \in T_2}[f(x,w) - 0]^2 \ \}. \qquad (C3)$$

For large $N$ and if the number of samples from each of the classes is in proportion to the *a priori* probability of class membership $P(T_j)$ this can be replaced by an integral

$$E = \int [f(x,w) - P(T_1|x)]^2 P(x)dx + P(T_1) - \int P^2(T_1|x)P(x)dx. \qquad (C4)$$

The minimum of this function with respect to $w$ clearly occurs for

$$f(x,w) = P(T_1|x), \qquad (C5)$$

so the output of the network can be interpreted as the *a posteriori* probability. This can be generalised for multiple output. It is reassuring (and should be used as a diagnostic) that the probabilities in an 'ideal' network add up approximately to unity. Moreover, if both the training and testing sets are drawn from the same parent distribution, then the frequency distribution $P(T_j)$ for the objects as classified by the ANN is similar to that of the training set. The link between minimum variance and probability also illustrates why a classification scheme where one calculates the Euclidean distance of the ANN output from the vector representing each of the possible classes and then assigns the object to the class producing the minimum distance is equivalent to assigning a class according to the highest probability (cf. Richard & Lippmann 1991). For a sigmoid output (eq. B3) it can be shown (Gish 1990) that the argument of the sigmoid, $z(x,w) = \ln[f(x,w)/(f(x,w) - 1)]$, with $f(x, w) = P(T_1|x)$ (eq. C4) and $P(T_2|x) = 1 - P(T_1|x)$ gives

$$z(x, w) = \ln \frac{P(T_1|x)}{P(T_2|x)}, \qquad (C6)$$

i.e. the argument of the sigmoid is modelling the log-likelihood ratio of the two classes. With the transfer function $\tanh(z) = 2/[1+\exp(-2z)] - 1$ the interpretation is similar. We note that the above analysis (eq. C4) does not tell anything about the network architecture, and it only holds for 'idealized' network and data. For more rigorous and general Bayesian approaches for modelling ANNs see MacKay (1992).

**C.2 Linear Networks and Wiener Filtering**

The weights, the free parameters of the ANN, can have a simple interpretation when the network is linear $[f(z) = z]$ without hidden layers, commonly called the 'perceptron'. For simplicity of notation we consider a network with a single continuous output, e.g. yielding the type $T$. In this case we can write the cost function as:

$$E = \frac{1}{2}\frac{1}{N}\sum_{\mu=1}^{N}[T^\mu - \sum_{k=0}^{M} w_k x_k^\mu]^2, \qquad (C7)$$

where $\mu = 1, ...N$ labels the objects, and $k = 0, ...M$ the parameters. The index $k = 0$ stands for the 'bias' term (with $x_0 = -1$), and it plays the role of an additive constant $w_0$ in the network equation. The minimum of $E$ with respect to the weights occurs at

$$\frac{\partial E}{\partial w_j} = \frac{1}{N}\sum_{\mu}[T^\mu - \sum_k w_k x_k^\mu]x_j^\mu = 0. \qquad (C8)$$

giving

$$\sum_k \langle x_k x_j \rangle w_k = \langle T x_j \rangle \qquad (C9)$$

where $\langle ... \rangle$ are averages over the $N$ objects.

The solution of this set of linear equations (for $j = 1, ...M$) for the optimal weights vector $\mathbf{w}$ can be written as

$$\mathbf{w}_{\text{opt}} = A^{-1}\,\mathbf{b}, \qquad (C10)$$

where $A_{jk} = \langle x_k x_j \rangle$ and $b_j = \langle T x_j \rangle$. More generally, if there are multiple output units (say a vector $\mathbf{s}$) so the weights form a matrix $W$ (not necessarily square), the minimum variance $\langle (\mathbf{s} - W\mathbf{x})(\mathbf{s} - W\mathbf{x})^T \rangle$ with respect to the weights occurs for

$$W_{\text{opt}} = \langle \mathbf{s}\mathbf{x}^T \rangle \langle \mathbf{x}\mathbf{x}^T \rangle^{-1}. \qquad (C11)$$

This is in fact the standard Wiener (1949) filter known in digital filtering and image processing, commonly applied for signal+noise problems with **x** = **s** + **n** (e.g. Rybicki & Press 1992 for a review, and Zaroubi et al. 1995 and references therein for recent cosmological applications). We note that the same result can be derived by conditional probabilities with Gaussian probability distribution functions, as well as by regularisation with a quadratic prior.

For an alternative, somewhat more complicated expression see Hertz et al. (1991, pg. 102), where the weights are given in terms of a covariance matrix of the *objects* (useful for the case of many features and few objects).

One can go one step further to generalize the above to non-linear *input*. This can be done e.g. by expanding the elements of the input vector as products of their powers. For example, if the input parameters are $x_1$ and $x_2$ the expanded input vector is

$$[1, x_1, x_2, x_1^2, x_1 x_2, x_2^2, ...].$$

This is sometime called the Volterra Connectionist model (VCM; see e.g. Rayner & Lynch 1989, Pao 1989, Lasenby and Fitzgerald 1993). Other alternatives for non-linear input are e.g. 'radial basis functions' and spherical harmonics. In fact, this can be viewed as an ad-hoc hidden layer which forces the input to a new non-linear form. The advantage of VCM network is that the global minimum is unique. This provides a reproducible solution and allows fast training. On the other hand, as the connections between input and 'hidden' layer are 'hard wired', the freedom of the network for difficult data sets is limited.

### C.3 Regularisation and Weight Decay

As in other inversion problems, the determination of many free parameters, the weights $w_i$'s in our case, might be unstable. It is therefore convenient to regularise the weights, e.g. by preventing them from growing too much. In the ANN literature this is called 'weight decay'. This approach is analogous to Maximum Entropy, and can be justified by Bayesian arguments, with the regularising function acting as the prior in the weight space. Note that this is a different application of Bayes theorem from the one discussed in §C.1, applied in the class-space.

One possibility is to add a quadratic prior and to minimize

$$E_{tot} = \alpha E_w + \beta E_D, \tag{C12}$$

where $E_D$ is our usual cost function, based on the data (e.g. eqs. B5 and C6) and

$$E_w = \frac{1}{2} \sum_{i=1}^{Q} w_i^2 \quad (C13)$$

is the chosen regularising function, where $Q$ is the total number of weights. The coefficients $\alpha$ and $\beta$ can be viewed as 'Lagrange multipliers'. While sometime they are specified ad-hoc, it is possible to evaluate them 'objectively' by Bayesian arguments in the weight-space. This has been done in the context of ANNs by MacKay (1992, see also Ripley 1993) following earlier analysis in relation with Maximum Entropy by Gull(1989; see also Lahav & Gull 1989). The Bayesian analysis gives the conditions on $\alpha$ and $\beta$ as

$$\chi_w^2 = 2\alpha \hat{E}_w = \gamma \quad (C14)$$

and

$$\chi_D^2 = 2\beta \hat{E}_D = N - \gamma \quad (C15)$$

where $N$ is the number of data points (objects) and

$$\gamma = \sum_{q=1}^{Q} \frac{\lambda_q}{\lambda_q + \alpha} \quad (C16)$$

where the $\lambda_q$'s are the eigen-values of the Hessian (in the weight-space) $\beta \nabla \nabla E_D$, evaluated with the weights at which $E_{tot}$ is minimum. The parameter $\gamma$ has an interesting interpretation, the number of 'well-determined' weights. If $\lambda_q \gg \alpha$ then $\gamma \approx Q$ (the total number of weights). In this case $\chi_D^2 \approx N - Q$, which is similar to the usual condition of $\chi^2$ equals the number of degrees of freedom. Moreover, if $Q \ll N$ then

$$E_{tot} \approx \frac{1}{\sigma_w^2} E_w + \frac{1}{\sigma_D^2} E_D \quad (C17)$$

where $\sigma_w^2 = 2\hat{E}_w/Q = \sum w_i^2/Q$ and $\sigma_D^2 = 2\hat{E}_D/N$, as expected for Gaussian probability distribution functions. We note that this analysis makes sense if the input and output are properly scaled e.g. between [0, 1] with sigmoid transfer functions, so all the weights are treated in the regularisation process on 'equal footing'. It can be generalized for several regularising functions, e.g. one per layer.

We note that the addition of the regularisation term $E_w$ changes the location of the minimum, now satisfying

$$\nabla E_D = -\frac{\alpha}{\beta}\nabla E_w = -\frac{\alpha}{\beta}\mathbf{w}, \qquad (C18)$$

as from eq. (C13) $\nabla E_w = \mathbf{w}$. The effect of the regularisation term here reminds the restoring force of harmonic oscillator: the larger $\mathbf{w}$ is the more it will get suppressed. The addition of the regularisation term to eq. (C4), gives a minimum for the extended cost function which does not satisfy eq. (C5), i.e. it violates the probabilistic interpretation in the class-space. However, one could construct a network with regularisation which will produce probabilities self-consistently (e.g. MacKay 1992). The weight decay term also modifies the Wiener solutions in §C.2

**FIGURE CAPTIONS**

**Figure 1a** The distribution of 5217 ESO-LV galaxies of all morphological types (top right) in the 2-dimensions defined by the first and second Principal Components as derived from PCA using 13 galaxy parameters. The other three panels show subsets of this fuzzy cloud according to their classification labels Sa+Sb, Sc+Sd and E+S0 as given in ESO-LV. The different morphological types occupy distinct regions in this new parameter space.

**Figure 1b** The top-right panel is as in Figure 1a, and the other 3 panels are for the classes E, S0, and Irr. Note that E and S0 galaxies are segregated.

**Figure 2** A schematic diagram of an encoder network with $M$ input parameters, $N$ (labeled as $M'$ in the text) nodes on the hidden layer, and $M$ output nodes. $N$ will range between 1 and $M$, depending on the desired data compression factor. During training, set Input=Output to teach the encoder to reproduce a given input vector at the output layer. This network performs PCA-like dimensionality reduction when the transfer function is linear, and can be extended to perform non-linear mapping.

**Figure 3** The cost function vs. the number of hidden units in encoder network with a sigmoid transfer function. The network was trained on the 5217 ESO-LV galaxies, each with 13 parameters. The cost function seems to drop roughly exponentially with the number of hidden units.

**Figure 4** An ANN configuration with $M$ input parameters, $N$ hidden nodes and a single 'analog' output. Such a network can perform a non-linear regression, and is used in our problem to predict the $T$-type, based on input galaxy parameters. All nodes in a given layer are connected to all nodes in the next layer. The 'bias' node allows additive constants to the network equation.

**Figure 5** The type $T_{net}$ predicted by the ANN for 1700 ESO-LV galaxies (based on a different set of 1700 galaxies) against the ESO-LV human classification $T_{eso}$. The Spearman correlation coefficient in this diagram is 0.83, and the average rms dispersion is 2.0 $T$-types.

**Figure 6** ANN reproducibility: A comparison between the predicted $T$-type of the network for 1700 ESO-LV galaxies from two runs, starting the minimization with different random weights. The Spearman correlation coefficient is 0.98 and the rms dispersion is 0.6 $T$-types. Note the transition between early to late type, discussed in the main text.